\documentclass[%
 reprint,
superscriptaddress,
 amsmath,amssymb,
 prl,
]{revtex4-1}

\usepackage{amssymb}
\usepackage{amsmath}
\usepackage{graphicx}
\usepackage{epsfig}
\usepackage{rotating}
\usepackage{verbatim}
\usepackage{wrapfig}
\usepackage{float}
\usepackage[export]{adjustbox}
\usepackage{subcaption}
\usepackage[skip=2pt,font=small]{caption}
\usepackage{ftnxtra}
\usepackage{physics}
\usepackage{color}
\usepackage{graphicx}
\graphicspath{{figures/}}
\usepackage{dcolumn}
\usepackage{bm}

\usepackage[
total={6.5in,9.75in}, top=0.9 in, left=0.9in, includefoot,
]{geometry}
\begin{document}

\preprint{APS/123-QED}

\title{Concentration banding instability of a sheared bacterial suspension}

\author{Laxminarsimharao V}
\affiliation{Engineering Mechanics Unit, Jawaharlal Nehru Centre for Advanced Scientific Research,
Bangalore 560064, India}
\author{Piyush Garg}
\affiliation{Engineering Mechanics Unit, Jawaharlal Nehru Centre for Advanced Scientific Research,
Bangalore 560064, India}
\author{Ganesh Subramanian}%
 \email{sganesh@jncasr.ac.in}
\affiliation{Engineering Mechanics Unit, Jawaharlal Nehru Centre for Advanced Scientific Research,
Bangalore 560064, India}


\begin{abstract}
We demonstrate a novel shear-induced mechanism for growth of concentration fluctuations in a bacterial suspension. Using a linear stability analysis, a homogeneously sheared suspension is shown to support exponentially growing layering perturbations in the shear-rate and bacterial concentration. Non-linear simulations show that the instability eventually leads to gradient-banded velocity profiles, with a local depletion of bacteria at the interface between the bands. Our results show that long-ranged hydrodynamic interactions are sufficient to explain recent observations of shear-bands in bacterial suspensions.  


%

\end{abstract}

\maketitle


Long-ranged hydrodynamic interactions in dilute bacterial suspensions drive growing orientation fluctuations, in turn leading to collective motion on length scales much larger than a single bacterium \cite{ganesh11, ganesh09, saintillan08, saintillan08b, simha02}. While large-scale coherent motion in unsheared bacterial suspensions observed in simulations \cite{saintillan07,graham08,ganesh15}, and in many experiments \cite{dunkel2013,wu2000,clement2014}, is regarded as well understood theoretically, much less is known about the dynamics of sheared bacterial suspensions \cite{clement16}. Several recent experiments have observed counter-intuitive behavior of bacterial suspensions under an external shear, including regimes of apparent superfluidity \cite{samanta18,clement16,lopez15,sokolov09,koch14b,stocker16}. In this letter, we demonstrate a novel concentration-shear coupled mechanism for growth of fluctuations in bacterial suspensions, eventually leading to banded steady states. The proposed mechanism is shown to lead to shear-bands, with concentration inhomogeneities, in the dilute regime itself; in sharp contrast to both passive complex fluids \cite{fardin16, cates06, olmsted08, fardin12, dhont08}, and active fluids \cite{ramaswamy13,fielding08,fielding11,liverpool10,liverpool18} where shear banding is observed/predicted only in the semi-dilute and concentrated regimes.

\begin{figure}
\center
\includegraphics[scale=.65]{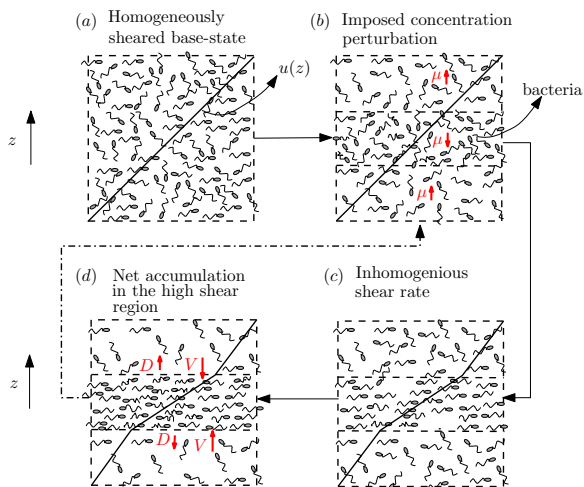}
\caption {Schematic illustrating the physical mechanism for growing concentration fluctuations in a sheared bacterial suspension. In this figure $\mu$, $V$ and $D$ represent the suspension viscosity, the destabilizing drift and the stabilizing diffusivity, respectively.}
\label{FIG11}
\end{figure}

Fig.~\ref{FIG11} illustrates the physical mechanism underlying the novel concentration-shear banding instability in a homogeneously sheared bacterial suspension. The bacteria are modeled as slender particles that swim along their axis, while being rotated and aligned by the background shear \cite{stocker14,ganesh17}. The latter leads to a spatially homogeneous suspension with an anisotropic orientation distribution (Fig.~\ref{FIG11} (a)). In the dilute regime, the contribution of the anisotropically oriented bacteria to the suspension viscosity is proportional to the local concentration. However, in sharp contrast to passive microstructural elements, the flow perturbation created by the tail-actuated swimming mechanism of bacteria (termed `pusher') aids the imposed shear, thereby lowering the suspension viscosity below that of the solvent  \cite{ganesh17, ramaswamy04, saintillan18,clement13,clement16,sokolov09,lopez15,haines09}. An initial gradient-aligned concentration perturbation thus leads to a lower (higher) effective suspension viscosity in regions of higher (lower) concentration. The invariance of the shear stress in the inertialess limit then implies that the higher (lower) concentration layers are subject to a higher (lower) shear rate (Fig.~\ref{FIG11} (c)).  In the higher shear rate region, the bacteria are more aligned with the flow. In turn, this implies a net concentration drift of bacteria into the higher shear rate (higher concentration) region, with a diffusivity driving an opposing stabilizing flux. The drift overcoming the diffusivity thus provides a mechanism for exponential growth of gradient-aligned (layering) concentration-shear fluctuations from the homogeneous state (Fig.~\ref{FIG11} (d)). Front-actuated swimmers (`pullers') such as algae, and passive rigid rods, increase the suspension viscosity in the dilute regime, leading to a stabilising drift, and thence, to decaying fluctuations.  

Migration of bacteria towards higher-shear rate regions, in inhomogeneous shear flows, leading to so-called shear trapping, has been examined before \cite{clement16, stocker14, stocker15, saintillan15, bearon15, sokolov2016}. However, all of these studies have focused on the kinematic point of view where changes in the bacterial concentration and orientation distribution, and the resulting changes in the bacterial stress, do not couple back to the flow. The mechanism outlined above illustrates, for the first time, how concentration and shear-rate fluctuations can be dynamically self-sustaining in bacterial suspensions. The exponentially growing layering perturbations eventually lead to a banded steady state, with the high shear band containing a (marginally) higher concentration of bacteria. In the rest of the letter, the aforementioned mechanism is first demonstrated through a linear stability analysis, followed by the results of non-linear simulations.

Gradient banding in sheared active fluids has been studied using phenomenological continuum equations with a bulk nematic or polar order \cite{fielding08,liverpool10,liverpool18,fielding11,ramaswamy13}. Since only the simplest terms allowed by symmetry are retained in these phenomenological equations, they do not describe the shear-induced migration observed in dilute bacterial suspensions \cite{clement16,stocker14,saintillan15,bearon15}. Consequently, the concentration banding instability reported here is also not described by the active fluid equations. Indeed, \cite{fielding08, fielding11, liverpool10, liverpool18} only report the shear-modified orientation instability already seen in unsheared active fluids \cite{simha02, ramaswamy13}. In the specific context of bacterial suspensions, an earlier effort only examined vorticity-aligned perturbations, and therefore did not find the novel concentration-shear instability analyzed here \cite{saintillan11}. To the best of our knowledge therefore, this letter is the first demonstration of a shear-induced mechanism for gradient banding in an active fluid.

At the microscale, a bacterium swims with a speed $U_b$, and the swimming direction ($\mathbf{p}$) decorrelates via both rotary diffusion (with diffusivity $D_r$) and tumbling (at a mean rate $\tau^{-1}$). Using $\tau, H, U_{\infty}$, where $U_{\infty}/H$ is the imposed shear-rate, as the time, length and velocity scales, respectively, the kinetic equation for the bacterium phase-space probability density, $\Omega(\mathbf{x},\mathbf{p},t)$ in the dilute limit is given by \cite{ganesh09}
\begin{eqnarray}
\frac{\partial \Omega}{\partial t} & + & \epsilon \mathbf{p}. \nabla_{\mathbf{x}}\Omega-D_r\tau \nabla^{2}_{p}\Omega+Pe \nabla_{p} \cdot (\dot{\mathbf{p}}\Omega) \nonumber \\
&+&[\Omega- \frac{1}{4\pi} \int d \mathbf{p}^{\prime} \Omega(\mathbf{p}^{\prime})]=0,
\label{EQ:NDGoveq}
\end{eqnarray}
where $\epsilon =  U_b\tau /H$ is the ratio of the bacterium run length to the imposed length scale and $Pe = U_{\infty}\tau/H$ denotes the relative importance of the shear-induced and intrinsic reorientation time scales. Approximating the bacteria as slender force-dipoles, the rotation due to the flow is given by the Jeffery's relation, $\dot{\mathbf{p}}=\mathbf{E} \cdot \mathbf{p}+\boldsymbol{\omega} \cdot \mathbf{p}-\mathbf{p}(\mathbf{E}: \mathbf{p} \mathbf{p})$, where $\mathbf{E}$ and $\boldsymbol{\omega}$ are the strain rate and vorticity tensors tensors, respectively, associated with the local linear flow \cite{Jeffery1922}. \eqref{EQ:NDGoveq} is coupled to the inertialess momentum and continuity equations
\begin{eqnarray}
Pe \nabla^{2} \textbf{u}&=&-\nabla \cdot \Sigma^{B},\nonumber \\
\nabla \cdot \mathbf{u}&=&0,
\label{EQ:NDEqOfMotion}
\end{eqnarray}
where we use the stress scale $\mu \tau^{-1}$. We now approximate $\Sigma^{B}$ by its active contribution alone which, in a continuum framework, is given in terms of the bacterium force-dipole density as $-\mathcal{A} \int d\mathbf{p} \Omega(\mathbf{p})(\mathbf{p}\mathbf{p}-I/3)$. The non-dimensional parameter $\mathcal{A}=C n_0 L^{2}U_{b}\tau$, termed the activity number here \cite{baskaran09}, is a measure of the bacterial force-dipole density, where $ L$ is the bacterium length, $n_0$ the number density and $\mathcal{C} $  the bacterial force dipole strength, with $\mathcal{C}>0$ for  `pushers' \cite{ganesh09,simha02,saintillan08,saintillan08b}. As will be seen below, $\mathcal{A}$ and $Pe$ delineate the unstable regions.   

\begin{figure}
\center
\includegraphics[scale=0.17]{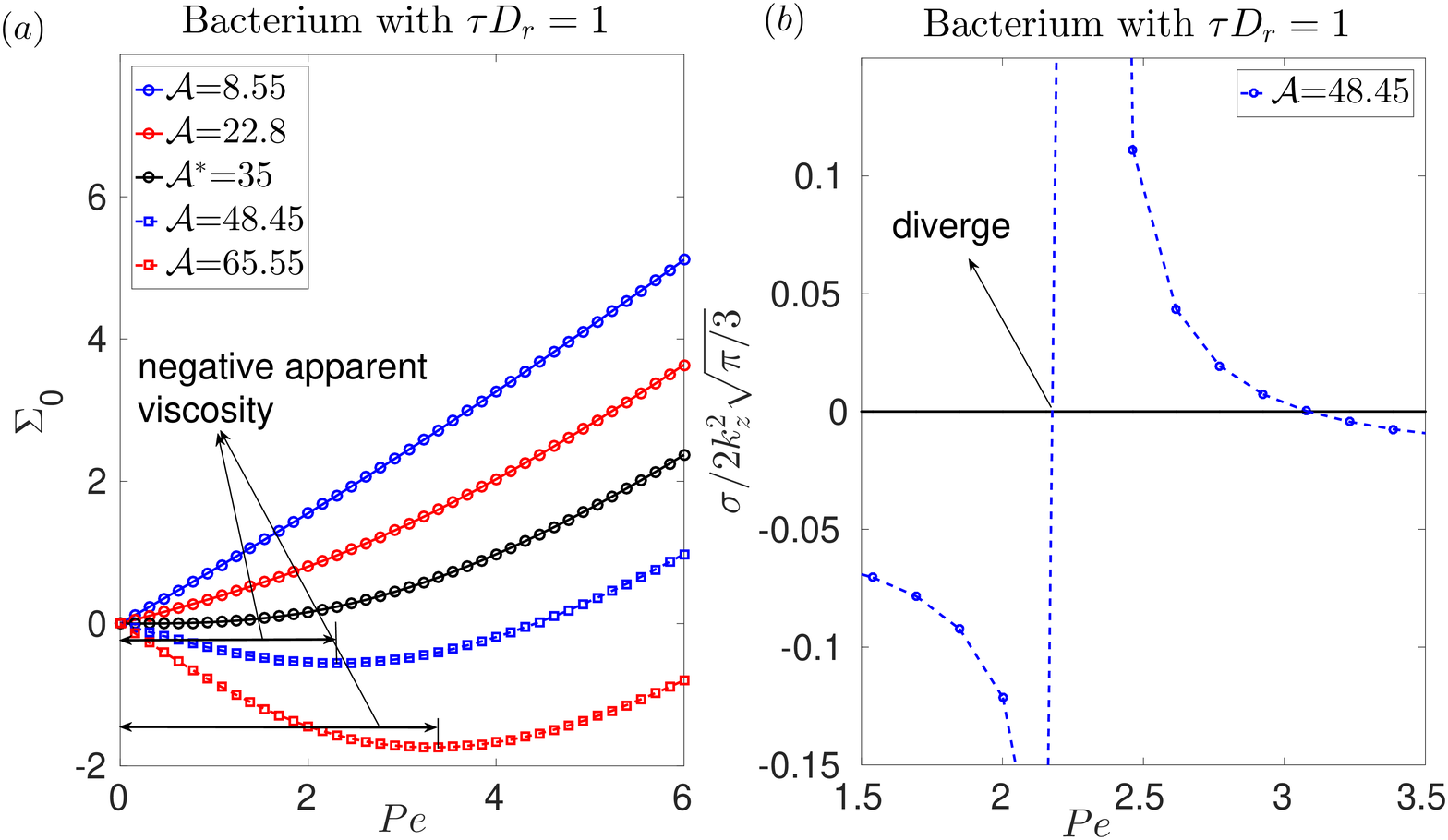}
\includegraphics[scale=0.17]{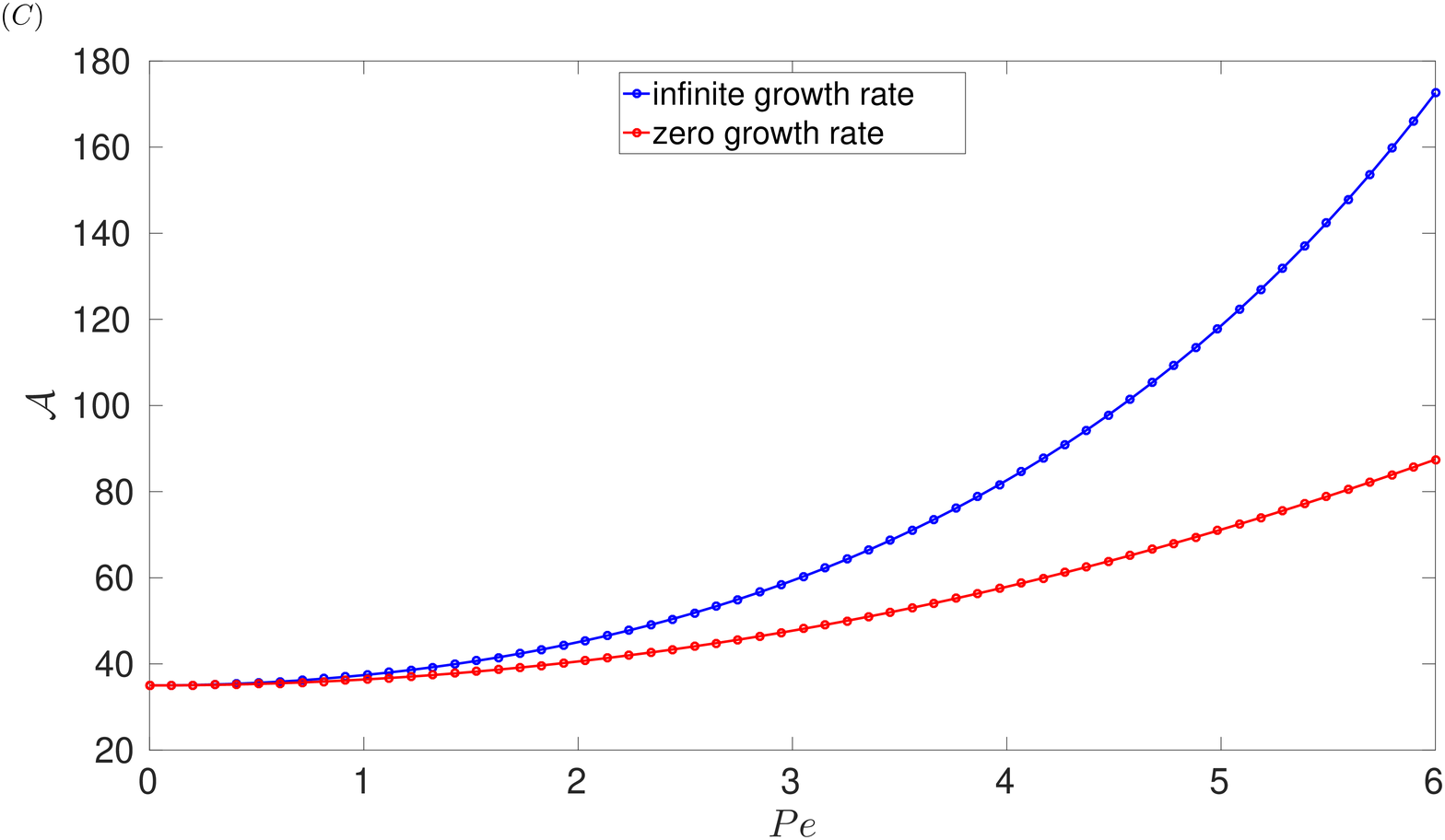}
\caption {Variation in the $(a)$ the homogeneous base-state stress and $(b)$ growth rate predicted by the multiple scales analysis against $Pe$. $(c)$ Unstable region in the $\mathcal{A}-Pe$ parameter plane.}
\label{FIG1}
\end{figure}

The homogeneous base-state is given by $u_{0}=z \mathbf{1_x}$ and an anisotropic orientation distribution {$\Omega_{0}(\mathbf{p})$, which needs to be solved for numerically \cite{supp}. Knowing $\Omega_0(\mathbf{p})$ allows the calculation of the stress-shear-rate curves for the homogeneous state; see Fig \ref{FIG1} (a) \cite{saintillan2010dilute,ganesh17,saintillan18,clement16}. For $\mathcal{A} <\mathcal{A}^{*}$, the base state stress ($\Sigma_{0}$) is a monotonically increasing function of the shear rate although the effective viscosity is lower than the solvent viscosity. $\mathcal{A}^{*} \approx 35$ marks the threshold for the instability in an unsheared suspension owing to the viscosity vanishing at $Pe=0$ \cite{ganesh09, ganesh11, saintillan08, saintillan08b, simha02}. For $\mathcal{A} >\mathcal{A}^{*}$, $\Sigma_{0}$ is a non-monotonic function of $Pe$ and the suspension has a zero viscosity at $Pe \equiv Pe_{cr}(\mathcal{A})$; $Pe_{cr}$ being an increasing function of $\mathcal{A}$. 

We examine the stability of the above homogeneous state to infinitesimal layering perturbations $(u_{1}$ and $\Omega_1)$ in the gradient direction \cite{note1}. Confinement is known to lead to concentration inhomogeneities via wall accumulation of swimming bacteria through both kinematic and hydrodynamic mechanisms \cite{saintillan15,baskaran17,lauga08,naji17,gompper16}. However, in order to focus on concentration inhomogeneities arising from banding in the bulk, we neglect wall effects in the analysis, and impose periodic boundary conditions in the non-linear simulations.

\textit{Concentration fluctuation dynamics} - In the limit $U_b\tau/H \rightarrow 0$, concentration fluctuations ($n_1=\int\Omega_1 d\mathbf{p}$) evolve on a slower, diffusive, time scale ($H^2/(\tau U^2_{b})$) compared to orientation fluctuations ($\tau$). A multiple scales analysis can thus be used to derive a generalized drift-diffusion equation for concentration fluctuations with the orientation fluctuations evolving in a quasi-static manner \cite{ganesh04, hinch97, koch12, koch14}. When linearized about the homogeneous base-state, we obtain \cite{supp} 
\begin{equation}
\frac{\partial n_1}{\partial t_{2}}= \frac{\partial}{\partial z} \left(-V_1+D_0 \frac{\partial n_1}{\partial z} \right),
\label{EQ:DriftDiffEq}
\end{equation}
with $V_1 = 2 \sqrt{\frac{\pi}{3}}  e_{1,0} \frac{\partial \dot{\gamma}_1}{\partial z}$. The perturbation shear-rate ($\dot{\gamma}_1$) is obtained from the momentum equation as
\begin{equation}
\frac{\partial \dot{\gamma}_1}{\partial z} =\mathcal{A} \frac{\partial n_1}{\partial z} \frac{\sqrt{\frac{2\pi}{15}}  \left(d_{2,-1}-d_{2,1} \right)}{\mu_0}.
\label{EQ:PertbMom}
\end{equation}
The constants involved in Eqs. (\ref{EQ:DriftDiffEq}) and (\ref{EQ:PertbMom}) are functions of $Pe$, and are obtained by numerically solving the linearized equations governing the quasi-static evolution of the orientation degrees of freedom \cite{supp}.

Assuming normal modes of the form $[n_1,\dot{\gamma}_{1}]=[\tilde{n}_1, \tilde{\dot{\gamma}}_{1}] \cos(zk_{z})\exp(\sigma t_{2})$, we obtain the following semi-analytical expression for the eigenvalue governing the evolution of concentration perturbations 
\begin{equation}
\sigma=k_{z}^2  \left(\mathcal{A} V_1 \frac{\sqrt{\frac{2\pi}{15}}  \left(d_{2,-1}-d_{2,1} \right)}{\mu_0}-D_{0} \right).
\label{EQ:GrowthRate}
\end{equation}  

The second term ($D_{0}$) in Eq.~\eqref{EQ:GrowthRate} represents the $Pe$-dependent stabilizing diffusivity. The first term represents the drift that drives a destabilizing flux from regions of low to high shear rate (Fig.~\ref{FIG11}), in proportion to the shear-rate gradient. When the drift exceeds the diffusivity, the homogeneous state becomes unstable (Fig~\ref{FIG1} (b)). 

For $Pe \rightarrow Pe_{cr}$, the suspension viscosity ($\mu_0$) vanishes and thus the destabilizing drift diverges in Eq. (\ref{EQ:GrowthRate}) making the suspension infinitely susceptible to concentration fluctuations (Fig~\ref{FIG1} (b)). The lower ($Pe_{cr}$) and upper ($Pe_{max}$) Peclet thresholds for the concentration-shear instability as a function of $\mathcal{A}$ are shown in Fig.~\ref{FIG1}~(c). The shear-rate range $(Pe_{cr}, Pe_{max})$ in which the system is susceptible to the concentration-shear instability increases with increasing $\mathcal{A}$. 

\textit{Coupled concentration and orientation fluctuation dynamics} - The divergence of the growth rate for $Pe \rightarrow Pe_{cr}$ is an artifact of the multiple scales analysis. For (dimensional) $\sigma \sim \tau^{-1}$ or $k_z \sim (U_b \tau)^{-1}$, the assumption of a separation of time scales between the concentration and orientation fluctuations is no longer valid. We therefore carry out a linear stability analysis, numerically, without the assumption of a time scale separation; Fig.~\ref{FIG2} (inset) shows good agreement between the two approaches. The full analysis continues to predict a finite growth rate of $\mathcal{O}(1/\tau)$ near $Pe_{cr}$.   

The multiple-scales analysis  does not predict a finite length scale for the fastest growing mode since $\sigma \propto k_z^2$ (see Eq~\ref{EQ:GrowthRate}). The full analysis, with orientation dynamics included, predicts the fastest growing wavenumber to be $\mathcal{O}(1/(U_b\tau))$ such that the relaxation times of the concentration and orientation (and thence, stress) fluctuations become comparable both being $\mathcal{O}(\tau)$ (Fig.~\ref{FIG3} (a)). For $k_z>\mathcal{O}(1/(U_b\tau))$, the diffusive rate of accumulation of bacteria ($k_z^{-2}/(\tau U^{2}_b)$) would exceed the stress relaxation time ($\tau$), and hence, such perturbations decay. For $Pe > Pe_{cr}$, there are strong, long wavelength concentration fluctuations ($\tilde{n}_{1}$) as predicted by the mechanism outlined earlier. This is seen in Fig.~\ref{FIG3} (b) where $\tilde{n}_{1}$ approaches a finite value even as $k_z \rightarrow 0 $ (with $\tilde{n}_{1} =0$ for $k_z =0 $ being a singular limit). Along with the multiple-scales analysis results, this reinforces the concentration-shear coupling mechanism that leads to a layering instability for $Pe> Pe_{cr}$.

\begin{figure}[!tbp]
  \centering
  \hfill
    \includegraphics[scale=0.16]{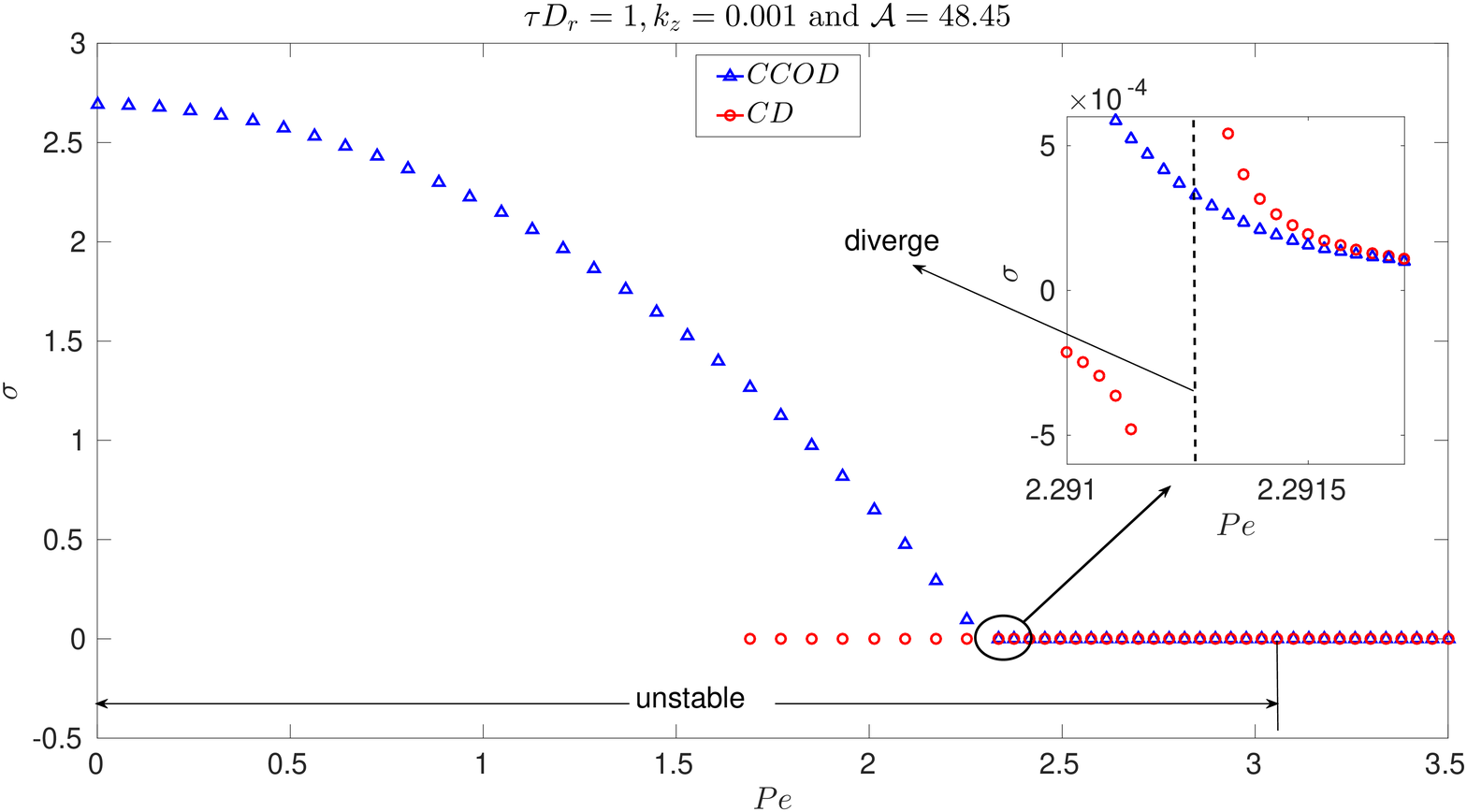}
\caption{Comparison of growth rates versus $Pe$ obtained from the concentration dynamics ($CD$) and coupled concentration-orientation dynamics ($CCOD$) analyses. The magnified inset emphasizes the agreement between the two approaches for $Pe > Pe_{cr}$}
\label{FIG2}
\end{figure}

The full analysis also predicts the orientation-shear instability, which has earlier been interpreted as a negative-viscosity instability responsible for the onset of collective motion in a quiescent bacterial suspension \cite{ganesh09, ganesh11, koch14}. Indeed, Fig.~\ref{FIG2} shows that orientation fluctuations drive an instability on the negative (effective) viscosity portion of the stress-shear-rate curve for $\mathcal{A} > \mathcal{A}^{*}$ and $Pe<Pe_{cr}$ where the multiple scales analysis predicts decaying concentration fluctuations (contrast with Fig.~\ref{FIG1}~(b) for $Pe < Pe_{cr}$). One therefore needs to distinguish between the orientation-shear and concentration-shear instability mechanisms which operate in distinct parameter regimes. The onset of instability coincides with the stress becoming a non-monotonic function of the shear rate (Fig~\ref{FIG1} (a)).
While the orientation-shear instability, analyzed by earlier authors \cite{ganesh09, saintillan08, saintillan08b}, is the usual mechanical shear-banding instability \cite{fardin16, cates06, olmsted08, fardin12, dhont08} operating in the range $Pe < Pe_{cr}$, the novel concentration-shear instability identified here exists only on the positive viscosity branch of the stress-shear curve.  

The physical mechanisms for the two instabilities can most easily be differentiated by focusing on the spatially homogeneous ($k_z=0$) mode. For $Pe < Pe_{cr}$, $k_z=0$ (implying no concentration fluctuations) is the fastest growing wavenumber with the growth driven by the orientation-shear coupling \cite{ganesh09, saintillan08b}. The growth rate of the $k_z=0$ mode monotonically decreases as $Pe$ increases. For $Pe > Pe_{cr}$, the dynamics is driven by concentration fluctuations and hence the $k_z=0$ mode is stable. 
In an unsheared suspension, the unstable eigenfunction does not have number density perturbations for any $k_z$ (Fig.~\ref{FIG3} (b)) in agreement with earlier predictions \cite{ganesh09, saintillan08b, saintillan11, shelley10}. Weak shear leads to weak long wavelength concentration fluctuations ($\tilde{n}_{1} \rightarrow 0$ as $k_z \rightarrow 0$) even for $Pe < Pe_{cr}$. However, as noted earlier, enhanced long wavelength fluctuations exist only for the concentration-shear instability ($Pe>Pe_{cr}$). For $Pe \sim Pe_{cr}$, there is no sharp distinction between the two mechanisms.

\begin{figure}
\center
\includegraphics[scale=.16]{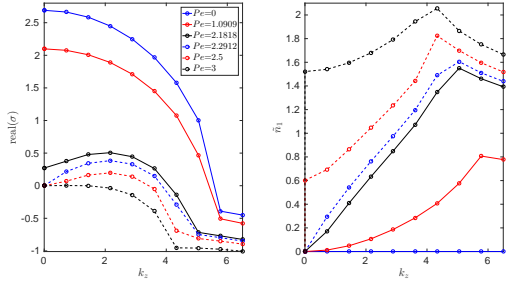}
\caption{Variation in the (a) growth rate and (b) concentration fluctuations ($\tilde{n}_{1}$) against the wavenumber for different shear rates with $\tau D_r=1$, $\mathcal{A}=48.5$ ($Pe_{cr} \approx 2.29$ and $Pe_{max} \approx 3.1$).}
\label{FIG3}
\end{figure}

\textit{Non-linear simulations} - To examine the steady state resulting from the linear instability discussed above, we numerically integrate \eqref{EQ:NDGoveq} and \eqref{EQ:NDEqOfMotion} in time. The non-linear simulations are carried out in two dimensions, so the orientation vector is restricted to the flow-gradient plane \cite{supp}. An imposed non-dimensional shear rate ($Pe$) is the control parameter.

Rather remarkably, the selected stress and shear-rate at steady state (see Fig~\ref{nonlinear}) can be explained using a Maxwell construction based on the homogeneous stess-shear rate profile. Fig~\ref{FIG1} (a) (with its symmetric extension for $Pe<0$) suggests a banded state with equal and opposite shear-rates ($\dot{\gamma}^{\star}$) with with a zero bulk stress and a homogeneous concentration \cite{note3}. In our numerical results, the selected stress (shear-rate) always differs from 0 ($\dot{\gamma}^{\star}$) by a finite amount but with a very small magnitude. With variation in the imposed shear rate, only the relative extents of the two bands change. The selected stress is, thus, (nearly) zero irrespective of $Pe$ and $\mathcal{A}$. Further, unexpectedly, the steady-state banded profiles do no show any major difference across $Pe_{cr}$ (Fig~\ref{nonlinear}) even though concentration fluctuations are crucial for the instability, and thus start-up kinetics for $Pe>Pe_{cr}$. This insensitivity of the selected stress to concentration-coupling is in sharp contrast to shear-banding in passive complex fluids, where it leads to an increase in the selected stress with the shear rate \cite{cates06, olmsted03, olmsted08}. 

\begin{figure}
\includegraphics[scale=0.85]{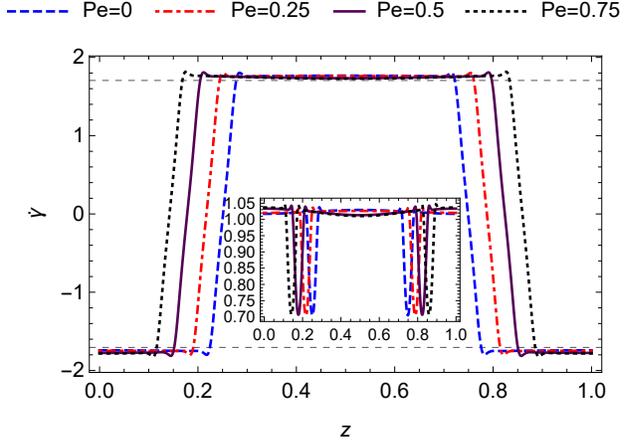}
\caption {The shear-rate ($\dot{\gamma}$) and concentration ($n$) (inset) profiles in the non-linear banded state for a box size 10 times the run length $U_b \tau$ for $\tau D_r=0.0025$ and $\mathcal{A}=62.83$ for which $Pe_{cr} \sim 0.67$. The shear rate ($\dot{\gamma}^{\star}$) is marked.}
\label{nonlinear}
\end{figure}

The equal and oppositely sheared zones in the banded state imply that the shear rate goes through zero at the interface, driving a local depletion of bacteria \cite{stocker14, bearon15, saintillan15} as seen in Fig~\ref{nonlinear}. Consequently, the bands have a marginally higher concentration of bacteria than the original homogeneous state, in turn implying that, in a finite domain, the shear rate selected slightly differs from $\dot{\gamma}^{\star}$ and that the stress is selected is finite, but (very) small in magnitude. The width of the interface between the shear bands is of the order of the bacterium run length $(U_b\tau)$, which can be seen from \eqref{EQ:NDGoveq} to be the length scale governing the spatial decay of stress \cite{olmsted08}. With increasing box size, the extent of interface depletion reduces, and the shear rate selected approaches  $\dot{\gamma}^{\star}$. 

An analogous result, for the selected stress, was obtained earlier for extensile active nematics for nematic-nematic banding and no concentration variation \cite{fielding08,fielding11}. The active-nematic formalism however has phenomenological constants that do not have a direct microscopic interpretation, especially for dilute bacterial suspensions that are far from an isotropic-nematic transition. Thus, \cite{liverpool18, liverpool10} report similar stress-shear rate curves and yet very different velocity profiles from those in \cite{fielding08, fielding11}. In contrast, our approach solves the underlying kinetic equation directly and rigorously demonstrates the selection of a banded-state even in the dilute regime. Crucially, our results demonstrate that long-range hydrodynamic interactions are sufficient to explain experimental observations of a banded state in dilute bacterial suspensions \cite{samanta18}. Postulating an orientationally ordered state, as is done in \cite{fielding08,fielding11, liverpool18, liverpool10}, is thus not necessary. 

\textit{Concluding Remarks} - In this letter, we have demonstrated a novel concentration-shear instability mechanism in dilute bacterial suspensions. The proposed instability is, in fact, reminiscent of the Helfand-Federickson mechanism that explains shear-enhanced concentration fluctuations in concentrated polymer solutions near an equilibrium critical point \cite{fredrickson89, onuki, milner93, larson92, leal13, leal14, olmsted03, pine91, pine92}. However, dilute bacterial suspensions are far from any critical point and the enhanced dynamics of the concentration fluctuations is crucially reliant on the novel rheological response arising from bacterial activity. We hope that the theoretical results reported in this letter would motivate light scattering experiments examining the dynamics of concentration fluctuations in bacterial suspensions. Similar experiments in polymer solutions have shed considerable light on on the nature of the shear-enhanced concentration fluctuations \cite{pine91,pine92}. 

The concentration-shear instability mechanism need not be restricted to a rheological scenario. Observations of collective motion driven by concentration fluctuations near the contact line of an evaporating drop were reported in \cite{koch14b}, and in pipe flow driven by a pressure gradient in \cite{stocker16}. The generalization of our results to an inhomogeneous shear-flow would lead to additional insight into these observations.       

\emph{Acknowledgements.} L.N.Rao would like to thank Science and Engineering Research Board, India (Grant No. PDF/2017/002050) and Jawaharlal Nehru Centre for Advanced Scientific Research, Bangalore for the financial support.

\bibliography{References}

\begin{thebibliography}{62}
\bibitem [{not({\natexlab{a}})}]{note1}%
  \BibitemOpen
  \href@noop {} {\bibfield  {journal} {\bibinfo  {journal} {Since
  gradient-aligned perturbations are the long-time limit for almost any initial
  wavevector in simple shear flow, this is not restrictive}\ }
  ({\natexlab{a}})}\BibitemShut {NoStop}%
\bibitem [{not({\natexlab{b}})}]{note3}%
  \BibitemOpen
  \href@noop {} {\bibfield  {journal} {\bibinfo  {journal} {Banded states with
  differing shear rates and concentrations and zero selected stress are also
  permitted by a Maxwell construction. However, we only observe such states
  transiently when starting from a finite amplitude concentration
  inhomogeneity. An external source of concentration inhomogeneity, such as no
  slip walls, could stabilise these states}\ } ({\natexlab{b}})}\BibitemShut
  {NoStop}%
\end{thebibliography}

\end{document}